\documentclass[twocolumn,prb,showpacs]{revtex4}

\usepackage{graphicx}
\usepackage{dcolumn}
\usepackage{bm}

\begin{document}

\title{Phase Diagram of the $t$--$U$--$V_1$--$V_2$ Model at Quarter Filling}

\author{Satoshi Ejima$^{(1)}$, Florian Gebhard$^{(1)}$, Satoshi
Nishimoto$^{(2)}$, and Yukinori Ohta$^{(3)}$}
\affiliation{$^{(1)}$Fachbereich Physik, Philipps-Universit\"at Marburg,
D-35032 Marburg, Germany} 

\affiliation{$^{(2)}$Institute for Theoretical Physics, University of G\"ottingen,
D-37077 G\"ottingen, Germany} 

\affiliation{$^{(3)}$Department of Physics, Chiba University,
Chiba 263-8522, Japan}


\begin{abstract}
We examine the ground-state properties of the one-dimensional 
Hubbard model at quarter filling with Coulomb interactions between
nearest-neighbors $V_1$ and next-nearest neighbors $V_2$.
Using the density-matrix renormalization group and exact 
diagonalization methods, we obtain for $U=10t$ three different phases
in the $V_1$-$V_2$ plane:
$2k_{\rm F}$- and $4k_{\rm F}$-charge-density-wave (CDW) and 
a broad metallic phase in-between. Assuming that the metal is 
a Tomonaga-Luttinger liquid (TLL), we calculate the TLL
parameter $K_{\rho}$. It is largest
when $V_1$ and $V_2$ are frustrated, and 
$K_{\rho}=0.25$ at the boundaries between the metallic 
phase and each of the two CDW phases.

\end{abstract}

\maketitle



The low-energy physics of one-dimensional correlated 
electron system can be described by the Tomonaga-Luttinger liquid
theory~\cite{Sch041}. In general,
the Green function and various correlation functions 
show a power-law behavior as a function of momentum $k$ and 
frequency $\omega$. The decay of the correlation functions is determined 
by the so-called TLL parameter $K_{\rho}$ which
depends on the strength of the interactions.
Numerous experiments as well as theoretical studies
have been performed on a variety of
systems in order to determine 
$K_{\rho}$ for one-dimensional metallic systems.

Most features of a recent angle-resolved photoemission spectroscopy (ARPES) 
experiment for the quasi one-dimensional organic conductor 
TTF-TCNQ~\cite{Cla02} can be explained by the photoemission 
spectral function of the one-dimensional Hubbard model with only on-site 
interaction~\cite{Ben04}. 
However, the Hubbard model yields 
$K_{\rho} \ge 0.5$, in contrast to $K_{\rho}\simeq 0.18$ as deduced from
angle-integrated data for the density of states~\cite{Cla02} which 
suggest a power-law behavior of the density of states near 
the Fermi level, 
$\rho(\omega) \propto \omega^\alpha$, 
where $\alpha=(K_{\rho}+K_{\rho}^{-1}-2)/4$ with $\alpha\simeq 1$.
Another example of a small $K_{\rho}$ is the 
quasi one-dimensional
organic conductor (TMTSF)$_2$X, where X=PF$_6$, ASF$_6$, 
or ClO$_4$. In this system, the parameter $K_{\rho}$ was 
estimated to be $K_{\rho} \simeq 0.23$ from the power-law dependence 
dominating in the higher-energy part of the optical conductivity, 
$\sigma(\omega) \sim \omega^{4a^2K_{\rho}-5}$, where $a$ is 
the order of the commensurability ($a=1$ for half filling; 
$a=2$ for quarter filling)~\cite{Eme79,Sol79}. This value is 
consistent with ARPES measurements~\cite{Zwi97}. 
TLL behavior was also suggested for the strongly anisotropic
transition-metal oxide PrBa$_2$Cu$_4$O$_8$; the parameter
was estimated to be 
$K_{\rho} \simeq 0.24$ from both the optical conductivity~\cite{Tak00} 
and the ARPES study of Zn-doped PrBa$_2$Cu$_4$O$_8$~\cite{Miz00,Miz02}.

Common features of all the materials mentioned above are that 
the filling of the conduction band is near one quarter,
$n \simeq 0.5$~\cite{Cla02,Sin03,Bou99,Tak00,Miz00}. Consequently,
they are close 
to a charge-density-wave (CDW) instability~\cite{Cla,Nad00,Cho00,Fuj03}.
In addition, the parameter $K_{\rho}$ is rather small, 
which is incompatible with the result from a simple one-dimensional
Hubbard model and shows
that the long-range Coulomb interactions 
between carriers is relevant. Therefore, we examine the 
one-dimensional Hubbard 
model with interactions to nearest and next-nearest neighbors,
\begin{eqnarray}
\hat H = &-& t \sum_{i,\sigma} (\hat c^\dagger_{i,\sigma}
\hat c_{i+1,\sigma} + {\rm h.c.}) + U \sum_i \hat n_{i,\uparrow}\hat n_{i,\downarrow}\nonumber\\
&+& V_1 \sum_{i,\sigma\sigma^\prime} 
\left(\hat n_{i,\sigma}-\frac{1}{2} \right) 
\left(\hat n_{i+1,\sigma^\prime}-\frac{1}{2} \right)\nonumber\\
&+& V_2 \sum_{i,\sigma\sigma^\prime} \left(\hat n_{i,\sigma}-\frac{1}{2} \right) 
\left(\hat n_{i+2,\sigma^\prime}-\frac{1}{2} \right)\; ,
\label{hamiltonian}
\end{eqnarray}
where $\hat c^\dagger_{i,\sigma}$ ($\hat c_{i,\sigma}$) is the creation 
(annihilation) operator of an electron with spin 
$\sigma (=\uparrow,\downarrow)$ at site $i$ and 
$\hat n_{i,\sigma}=\hat c^\dagger_{i,\sigma}\hat c_{i,\sigma}$ is the 
number operator.  $t$ is the hopping integral between 
neighboring sites, $U$ is the strength of the Hubbard interaction, and 
the charge frustrating interactions $V_1$ and $V_2$ 
determine the nearest-neighbor 
and next-nearest-neighbor Coulomb repulsion.
We restrict ourselves to the case of quarter filling ($n=1/2$),
and strong coupling, $U=10t$.

\begin{figure}[htbp]
  \begin{center}
    \resizebox{7cm}{!}{\includegraphics{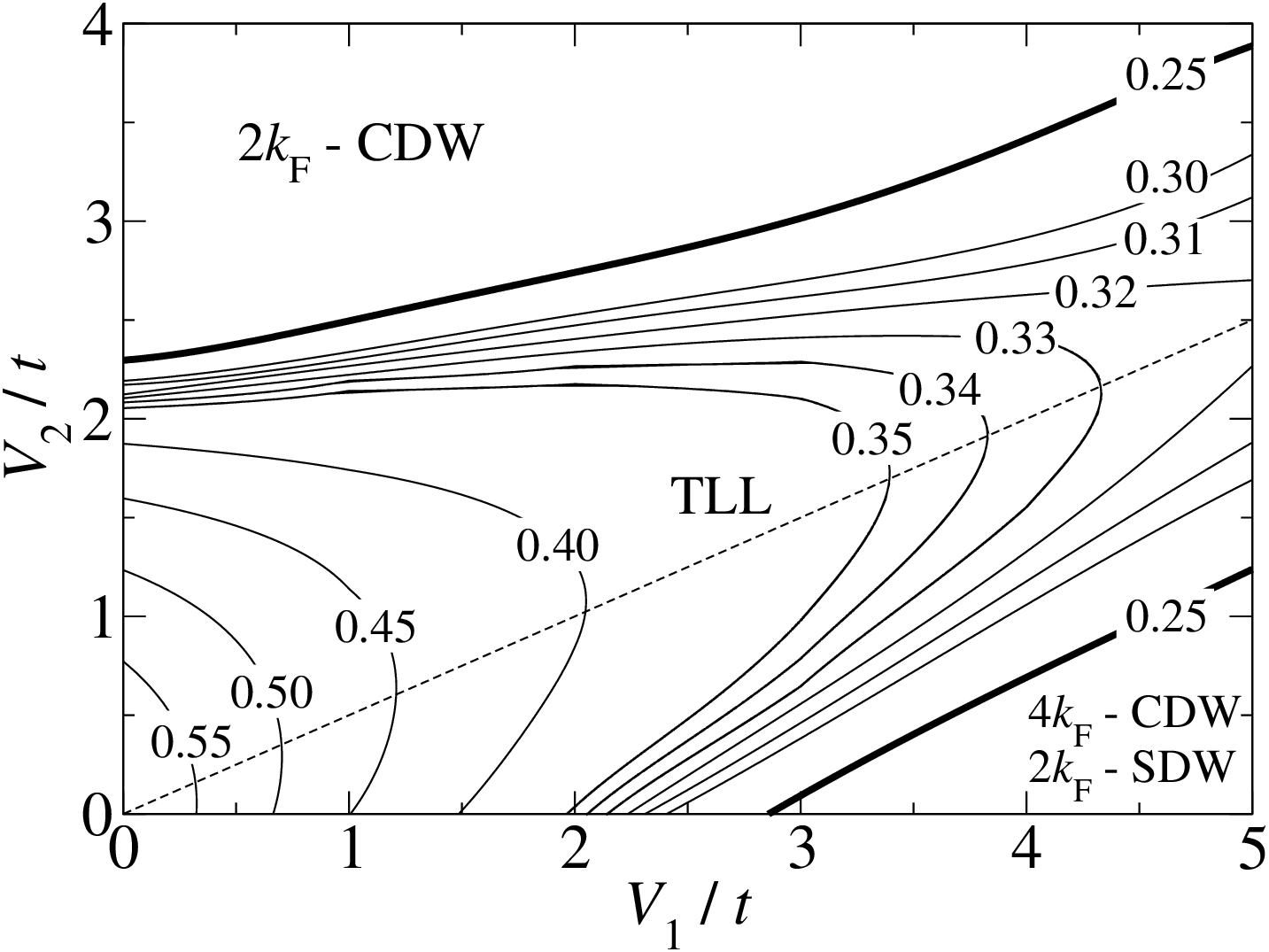}}
  \end{center}
  \caption{Phase diagram of the one-dimensional
$t$--$U$--$V_1$--$V_2$ model for $U=10t$ at 
quarter filling. The curves show the contours of constant 
TLL parameter $K_{\rho}$. Bold lines represent the 
boundary of the metal-insulator (CDW) transition and the dotted 
line corresponds to $V_2/t=V_1/(2t)$.\label{fig1}}
\end{figure}

So far several theoretical studies have been made on this 
and similar models~\cite{Dag97,Yos01,Seo01,Nis03,Zhu97,Sch04}. 
In the ground state there exist two CDW phases with wavenumbers 
$q=2k_{\rm F}$ and $4k_{\rm F}$. Between the two CDW phases 
there appears a wide region of a vanishing charge gap which 
results from the geometrical frustration of the long-range Coulomb 
interactions.
If we assumed the inter-site Coulomb repulsions to 
be inversely proportional to the inter-site distance, we would find
$V_1 = 2V_2$, so that neither of the two CDW instabilities dominates
in the atomic limit, $t=0$. Hence, 
one can easily imagine that 
the phase diagram contains a metallic region as soon as a finite~$t$ is 
introduced. However, little is known about the physical properties of 
this metallic state. 
It is the purpose of this paper to obtain an 
accurate ground-state phase diagram of the $t$--$U$--$V_1$--$V_2$ model, 
including the TLL parameter $K_{\rho}$ whereby we make the natural assumption
that the metallic phase is a Tomonaga-Luttinger liquid. 
Our phase diagram is shown in 
Fig.~\ref{fig1}.


We apply the density-matrix renormalization group (DMRG) method 
for the calculation of the ground-state energy. 
We study chains with up to 256 sites with open boundary 
conditions and keep up to $m=2000$ density-matrix
eigenstates so that the maximum truncation error is about $10^{-5}$.
We use periodic boundary conditions for the
calculation of charge excitations with
the Lanczos exact diagonalization technique.

\begin{figure}[thbp]
    \includegraphics[width=6.5cm]{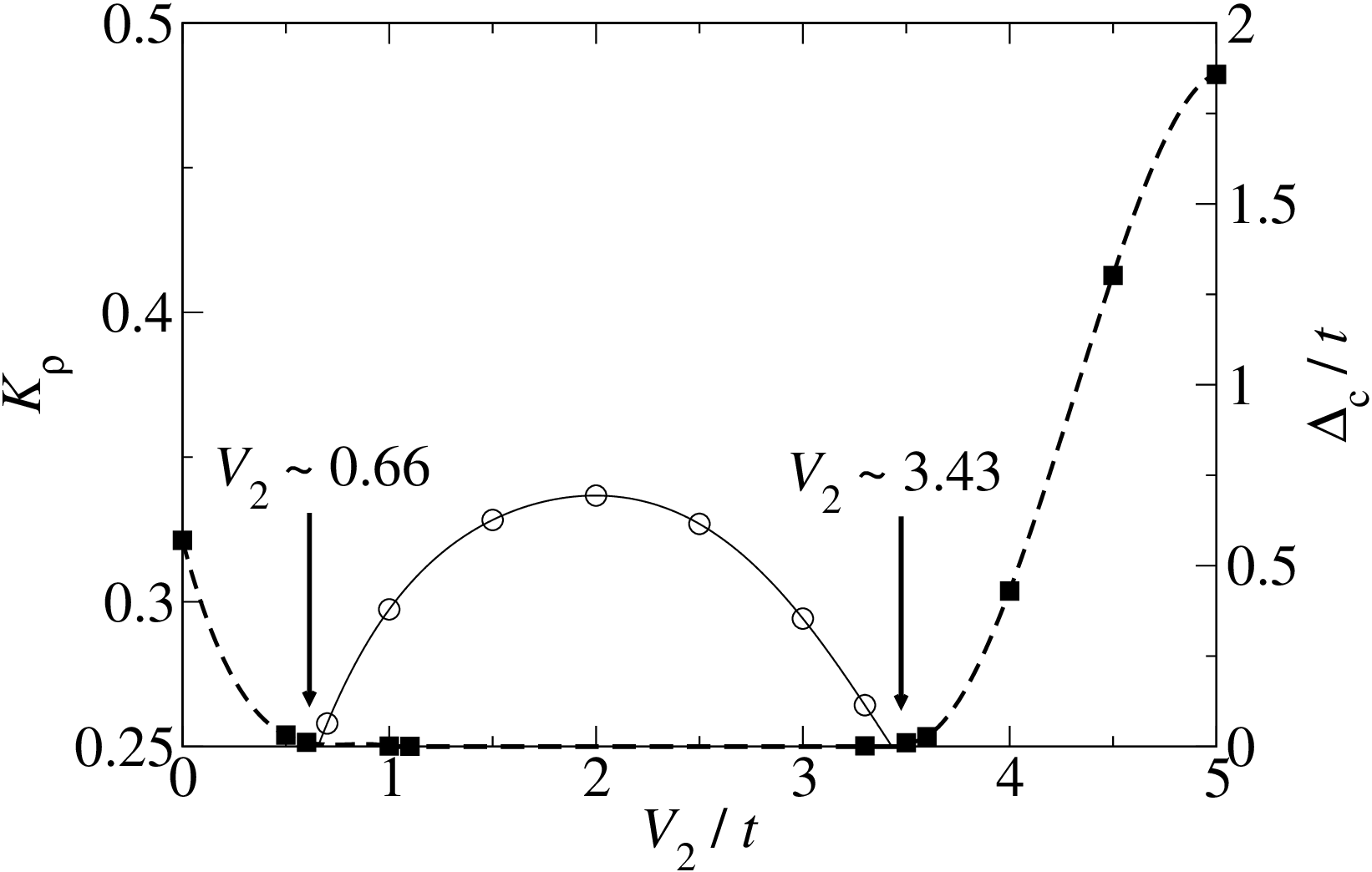}
  \caption{Charge gap~$\Delta_{\rm c}^{(2)}/(2t)$ (squares) 
and TLL parameter~$K_{\rho}$ (circles) in the one-dimensional
$t$--$U$--$V_1$--$V_2$ model at quarter filling 
for $U/t=10$ and $V_1/t=4$. Lines are guides to the eye.\label{fig2}}
\end{figure}

In order to determine the metallic region in the phase diagram,
we calculate the two-particle charge gap, 
\begin{eqnarray}
\Delta_{\rm c}^{(2)} (L) &=& 
 E_0 (N_{\uparrow}+1,N_{\downarrow}+1,L) 
\label{chargegap}
\\
&&+ E_0 (N_{\uparrow}-1,N_{\downarrow}-1,L) 
- 2E_0 (N_{\uparrow},N_{\downarrow},L)\; ,
\nonumber
\end{eqnarray}
where $E_0(N_{\uparrow},N_{\downarrow},L)$ denotes the 
ground-state energy of a chain of length $L$ with 
$N_{\uparrow}$ spin-up electrons and $N_{\downarrow}$ 
spin-down electrons. When pairing is absent, in the thermodynamic limit
the two-particle
charge gap becomes twice the single-particle charge gap,
$\lim_{L\to\infty} \Delta_{\rm c}^{(2)} (L)
=2 \lim_{L\to\infty}\Delta_{\rm c}(L)$.
We find that $\Delta_{\rm c}^{(2)}(L)$ decreases 
monotonically with increasing $L$, so that we can extrapolate it
to the thermodynamic limit systematically by 
performing a polynomial fitting in $1/L$. Note that we need  
relatively large system sizes to obtain an accurate 
phase boundary. The extrapolated results 
$\Delta_{\rm c}^{(2)}/t$ at $U/t=10$ and 
$V_1/t=4$ are shown in Fig.~\ref{fig2}. 

It is evident that $\Delta_{\rm c}^{(2)}/t$ is finite for both small 
$V_2/t \le 0.66$ and large $V_2/t \ge 3.43$, i.e., 
the system is insulating, and $\Delta_{\rm c}^{(2)}/t$ vanishes 
in a wide range of $V_2/t$, i.e., $0.66 \le V_2/t \le 3.43$,  
within the accuracy of the extrapolation (error smaller than $10^{-4}t$). 
We find similar situations for other values of $V_1/t$ as well, 
which demonstrates that a stable metallic phase indeed exists 
between two insulating phases as shown in Fig.~\ref{fig1}.
This result is consistent with other 
studies~\cite{Dag97,Seo01,Nis03,Zhu97,Sch04}.

\begin{figure}[ht]
\includegraphics[width=6.5cm]{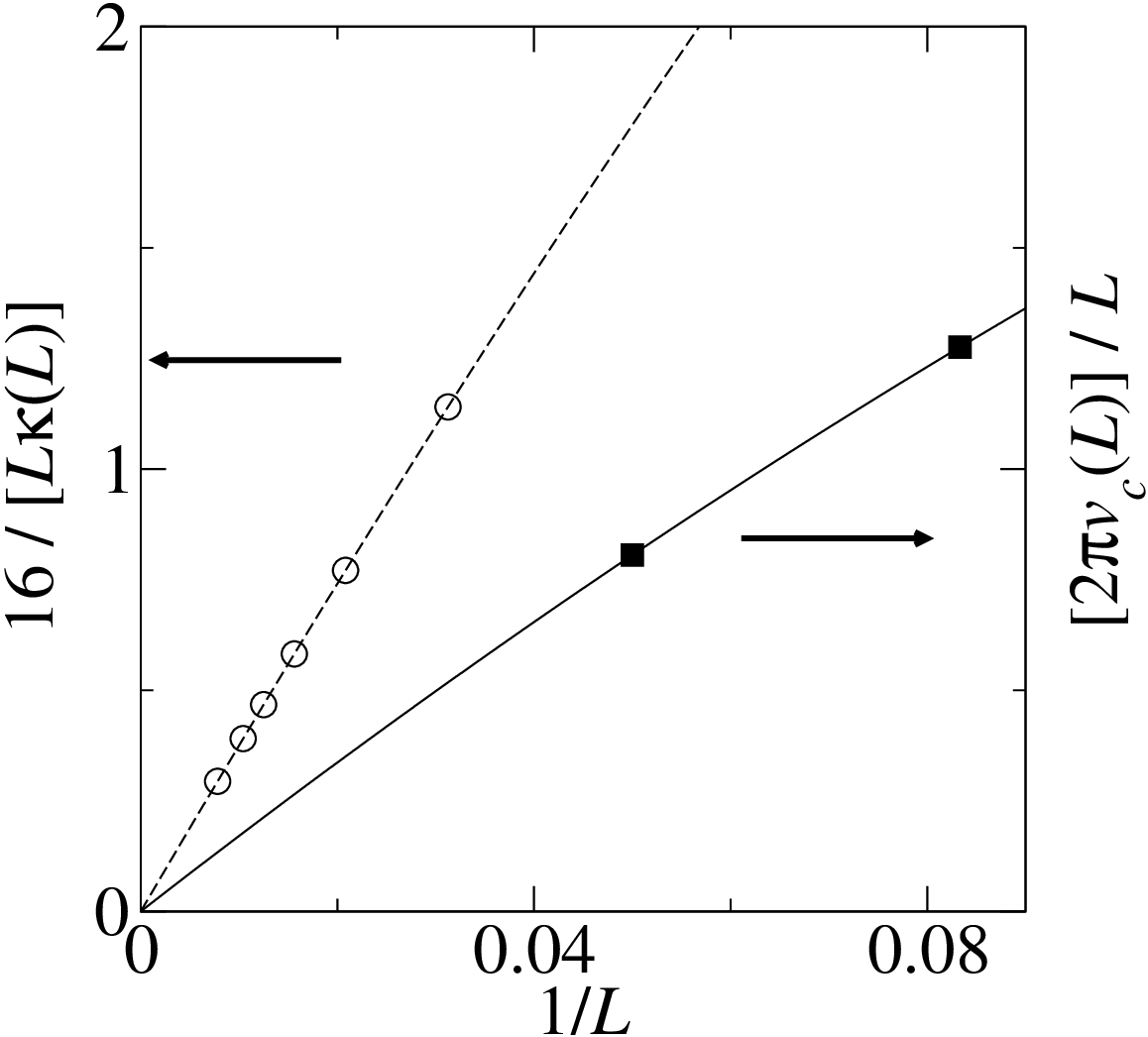}
\caption{Scaling of the ground-state 
energy differences from Eq.~(\ref{kappaL}) 
and the excitation energy from Eq.~(\ref{vcL}) as a function of 
$1/L$ at $U/t=10$, $V_1/t=1$ and $V_2/t=1$. The lines indicate 
the polynomial fits by Eq.~(\ref{polynomial}) to the data.\label{fig3}}
\end{figure}

In the Tomonaga-Luttinger phase, the dominant correlations are determined 
by a single parameter $K_{\rho}$, which is given by~\cite{Sch90,Kaw90,Fra90}
\begin{equation}
K_{\rho} = (\pi/2) n^2 \kappa v_{\rm c},
\label{Krho}
\end{equation}
where $\kappa$ and $v_{\rm c}$ are the compressibility and the 
charge velocity, respectively. 
To estimate $K_{\rho}$ numerically, we calculate the compressibility $\kappa$
from the charge gap, 
\begin{equation}
\Delta_{\rm c}^{(2)}(L)=4/(n^2 L\kappa(L))\;,
\label{kappaL}
\end{equation}
and the velocity of the charge excitations $v_{\rm c}$ from
\begin{equation}
E_{1 \rho}(N_{\uparrow},N_{\downarrow},L)
-E_0(N_{\uparrow},N_{\downarrow},L)=2\pi v_{\rm c}(L)/L\;,
\label{vcL}
\end{equation}
where $E_{1 \rho}(N_{\uparrow},N_{\downarrow},L)$ is the lowest 
excitation energy with momentum $k=2\pi/L$ and total spin $S=0$
for finite system size $L$. As shown in Fig.~\ref{fig3}, 
both of the quantities are monotonous functions of 
$1/L$ for all parameter sets and can be fitted to
a polynomial function 
\begin{equation}
f(L) = a_1 L^{-1} + a_2 L^{-2} + \cdots \; . 
\label{polynomial}
\end{equation}
Therefore, we obtain 
\begin{eqnarray}
\kappa&=&\lim_{L \to \infty} \kappa(L)=16/a_1^{(\kappa)} \; ,
\label{kappathermo} \\
v_{\rm c}&=&\lim_{L \to \infty}v_{\rm c}(L)=a_1^{(v_{\rm c})}/2\pi \; , 
\label{vcthermo}
\end{eqnarray}
so that the TLL parameter can be estimated as 
\begin{eqnarray}
K_{\rho}=a_1^{(v_{\rm c})}/a_1^{(\kappa)} 
\label{krhothermo}
\end{eqnarray}
in the thermodynamic limit. 
For the compressibility we can use DMRG for up to 128~sites.
The extrapolation is very well behaved, and a reliable extrapolation 
could have been obtained from the results for much smaller systems.
This makes us confident that the extrapolation for the charge velocity
is meaningful despite the fact that exact diagonalization is limited 
to $L\leq 20$.
A comparison with the exact $K_{\rho}^{\rm exact}$ 
for the one-dimensional Hubbard model ($V_1=V_2=0$)~\cite{Sch90}
shows that relative errors 
$|K_{\rho}^{\rm DMRG}-K_{\rho}^{\rm exact}|/K_{\rho}^{\rm exact}$
are below $1\%$.
For finite $V_1$ and/or $V_2$, our results are consistent with 
other estimates using the charge structure factor.~\cite{Dau98,Cla99,Eji04}

In Fig.~\ref{fig2}, the parameter $K_{\rho}$ is plotted as a function 
of $V_2/t$ at fixed $U/t=10$ and $V_1/t=4$. We see that $K_{\rho}>0.25$ 
when the charge gap is zero and $K_{\rho}=0.25$ at the critical points.
We have confirmed numerically that $K_{\rho}$ is always 0.25 at the
CDW critical points 
for all finite values of~$U$. $K_{\rho}$ reaches its maximum value around 
$V_1=2V_2$, i.e., the density-density correlations 
decay most rapidly when 
$V_1$ and $V_2$ maximally frustrate each other.
In general, long-range Coulomb repulsions are expected to suppress
the value of $K_{\rho}$. This is consistent with our results because 
$K_{\rho}$ decreases when the values of $V_1$ and $V_2$ deviate 
from the relation $V_1\approx 2V_2$, whereby the effective interaction 
strength increases. Apparently, the line $V_1=2V_2$ goes along 
the ridges of the contour line of $K_{\rho}$ in Fig.~\ref{fig1}. 
This has been already suggested in the spinless fermion case and 
similar models~\cite{Nis03,Sch04}.

\begin{figure}[htbp]
\includegraphics[width=7.5cm]{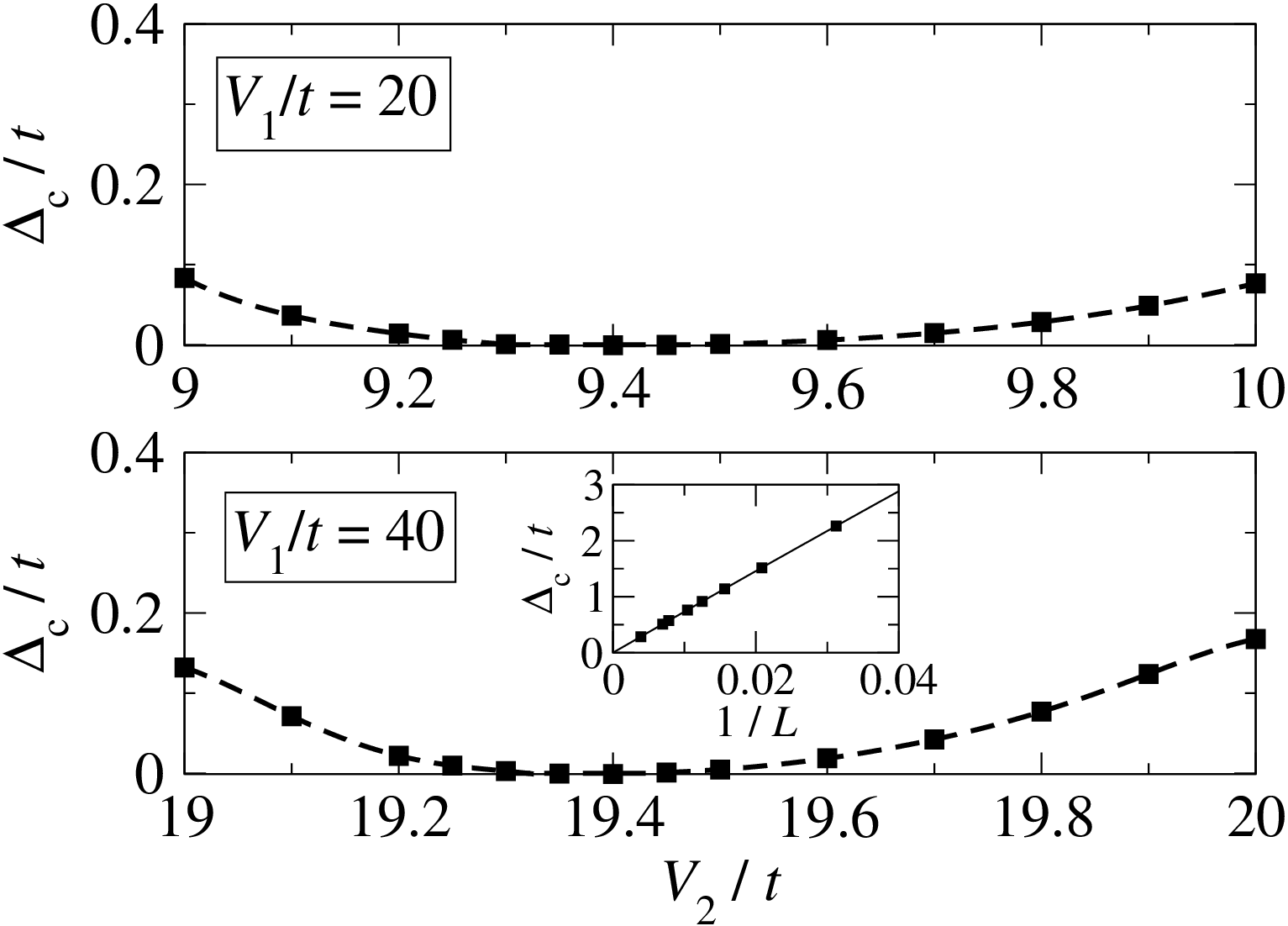}
\caption{Charge gap for spinless fermions at $V_1/t=20$ (upper part)
and $V_1/t=40$ (lower part). The charge gap vanishes around $V_2/t=
V_1/(2t)-0.6$ in both cases. The inset shows the charge gap for
$V_2/t=19.4$ as a function of $1/L$ for system sizes up to 
$L=256$.\label{fig4a}}
\end{figure}

In the limit of large values of $V_1$ and $V_2$ ($V_1,V_2<U$), 
the boundary between the ordered phases and the metallic phase 
shrinks.
In Fig.~\ref{fig4a}, we show the charge gap as a function of 
$V_2/t$ at $V_1/t=20$ and $V_1/t=40$ for spinless fermions ($U=\infty$). It 
demonstrates that a narrow but stable metallic regime exists between 
the two insulating phases. For large $V_1$ and $V_2$ it is very
difficult to extrapolate~$\Delta_{\rm c}^{(2)}$ reliably
from exact diagonalization data ($L\leq 20$), and
DMRG ($L = {\cal O}(200)$) must be used in the extrapolation.
For example,
Fig.~\ref{fig4a} shows that 
$V_1/t=40$ at $V_2/t=20$ is a CDW insulator whereas
it has been assigned a `non-Tomonaga-Luttinger metal' 
in Ref.~\onlinecite{Zhu97}.
Even with DMRG it is very difficult to decide whether or not
the metallic region shrinks to zero at some tri-critical point, 
as proposed in Ref.~\onlinecite{Sch04}.
If it exists it is far beyond the values quoted previously~\cite{Sch04}.
As seen from Fig.~\ref{fig4a} there is a metallic region for
$V_1/t=40$ in the vicinity of $V_2/t=19.4$. In the surrounding of this
point, the gap nicely scales to zero as a function of inverse system
size, as seen from the inset to Fig.~\ref{fig4a}.
Our results indicate that the metallic phase appears below the line 
$V_2=V_1/2$, 
around $V_2=V_1/2-0.6t$. We speculate that there is a metallic phase between 
the two CDW phases for all finite $V_1$ and $V_2$, 
i.e., there is no tri-critical points in the $V_1$-$V_2$ phase diagram.

In order to study the spin degrees of freedom,
we calculate the spin gap for system size $L$, defined by
\begin{eqnarray}
\Delta_{\rm s}(L)=E_0(N_{\uparrow}+1,N_{\downarrow}-1,L)
-E_0(N_{\uparrow},N_{\downarrow},L).
\label{spingap}
\end{eqnarray}
As in the case of charge gap, the extrapolation 
to the thermodynamic limit is straightforward.
In the $2k_{\rm F}$-CDW phase, 
we find that $\Delta_{\rm s}$ is always finite because the system contains
separated spin singlet pairs. For fixed
$V_1$, $\Delta_{\rm s}$ increases 
as a function of $V_2$ and eventually saturates at 
$\Delta_{\rm s}=4t^2/(U-V_1)$ in the limit $V_2 \to \infty$. 
This is readily understood because, for $U, V_2 \gg V_1$, 
the system can be mapped to an effective spin 
Hamiltonian 
$H=J_{2k_{\rm F}} \sum_i \hat{\bf S}_{4i} \cdot \hat{\bf S}_{4i+1}$ 
with $J_{2k_{\rm F}}=4t^2/(U-V_1)$. This model is trivial and  
the spin gap is the energy difference between the singlet and triplet
state at each bond, $\Delta_{\rm s}=4t^2/(U-V_1)$. 

In the $4k_{\rm F}$-CDW phase, we find that $\Delta_{\rm s}$ 
is always zero, because a charged site and a vacant site come 
alternately and the spin degrees of freedom can be described in terms
of a  one-dimensional uniform 
Heisenberg model. In fact, for $U > V_1/2 \gg V_2$, 
the effective spin Hamiltonian 
can be written as  
$H=J_{4k_{\rm F}} \sum_i \hat{\bf S}_{2i} \cdot \hat{\bf S}_{2i+2}$ 
with $J_{4k_{\rm F}}=4\left(t^2/(V_1-2V_2)\right)^2
\left[1/(U-V_2)+2/(U-2V_2)\right]$. This effective Heisenberg model displays
gapless spin excitations, in agreement 
with our numerical results. 

\begin{figure}[thbp]
    \includegraphics[width=6.5cm]{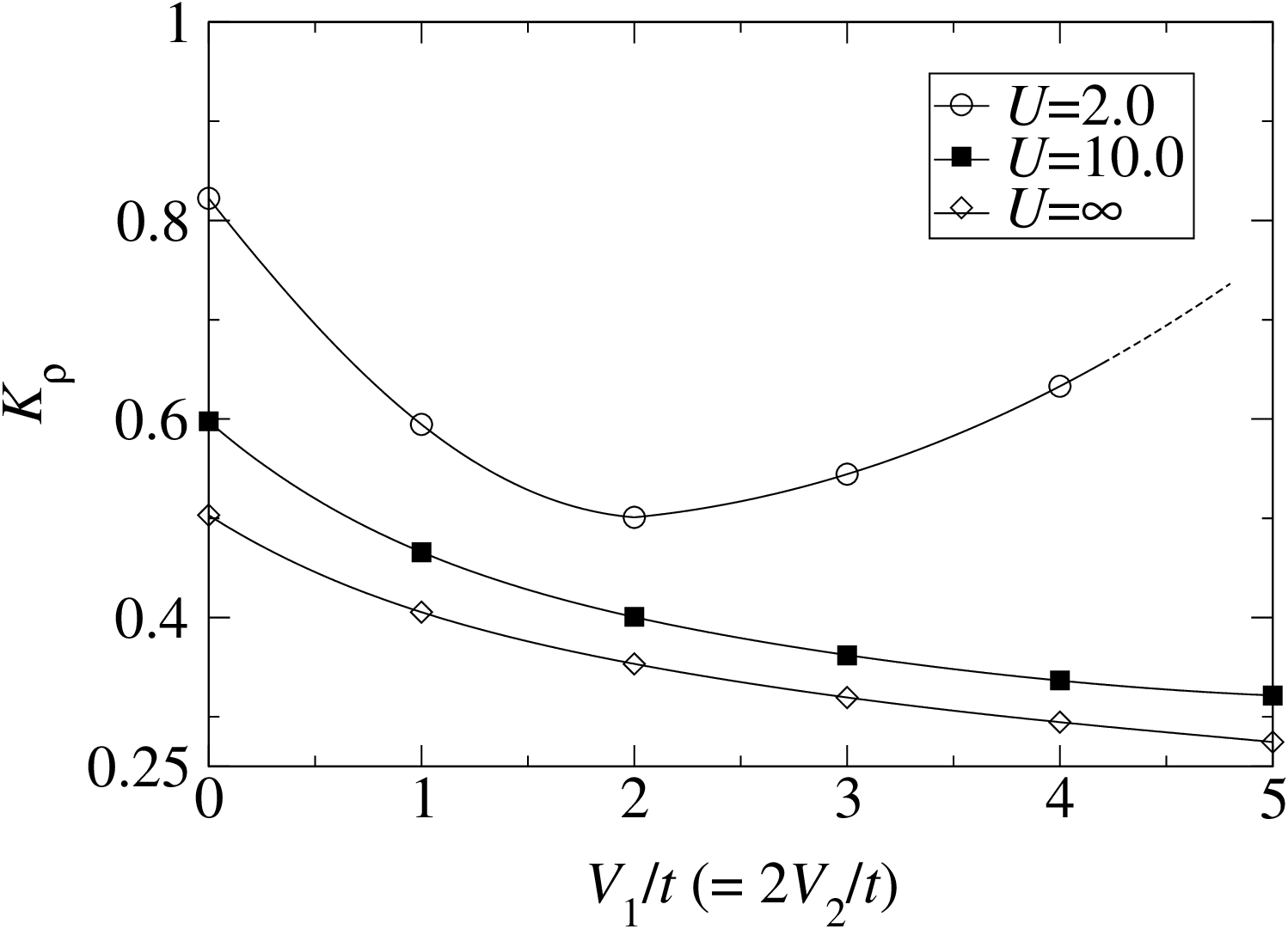}
  \caption{TLL parameter $K_{\rho}$ 
of the $t$--$U$--$V_1$--$V_2$ model at quarter filling 
as a function of $V_1/t$ along a line $V_1=2V_2$ for $U=2t$ 
(circles), $U=10t$ (squares) and $U=\infty$ (diamonds). The lines are 
guides to the eye.\label{fig4}}
\end{figure}

Lastly, in Fig.~\ref{fig4}, we plot the TLL parameter
$K_{\rho}$ as a function of $V_1/t$ for fixed $V_1=2V_2$
and several values of~$U$. For large~$U$, $K_{\rho}$ 
decreases as a function of $V_1/t$ and eventually crosses 
$K_{\rho}=0.25$ at some finite value of $V_{1,{\rm c}}$. As shown above,
$K_{\rho}\geq 0.25$ for a metallic phase, so that 
the Tomonaga-Luttinger liquid
turns into the $4k_{\rm F}$-CDW at $V_{1,{\rm c}}$.
For small~$U$, e.g., $U=2t$ in Fig.~\ref{fig4}, 
$K_{\rho}$ decreases as a function of $V_1$, displays a 
minimum around $V_1/t \approx {\cal O}(U/t)$ with $K_{\rho,{\rm min}}>0.25$,
and increases again. This results from the fact that $V_1$ overcomes
the Hubbard interaction~$U$ and electrons with opposite spin gain 
energy from on-site pairing. Eventually, 
$K_{\rho}$ can become larger than unity and superconducting correlations 
are dominant. 
As seen from Fig.~\ref{fig4}, the ground-state phase diagram
in Fig.~\ref{fig1} is representative 
for all~$U\gtrsim 4t$ when a superconducting phase does not interfere.
The ridge of $K_{\rho}$ goes along the line $V_1=2V_2$ and $K_{\rho}$ 
decreases
as the value of $V_1$ and $V_2$ deviate from this line.

In conclusion, we obtained an accurate ground-state phase diagram of 
the $t$--$U$--$V_1$--$V_2$ model at quarter filling using
DMRG and exact diagonalization methods. For intermediate to large
Hubbard interaction~$U\gtrsim 4t$, the system has 
CDW phases with $q=2k_{\rm F}$ and $4k_{\rm F}$ 
between which there appears a broad region of 
a Tomonaga-Luttinger liquid. Because 
of the geometrical frustration 
of the long-range Coulomb interactions, $K_{\rho}$ is maximum 
around $V_2=V_1/2$. It is smallest at the 
phase boundaries, $K_{\rho}=0.25$, which appears to be 
universal for transitions between the
metallic and the CDW phases~\cite{Sch04}.

\acknowledgments
We are grateful to E.~Jeckelmann, G.~Japaridze, and P.~Schmitteckert 
for helpful discussions. We acknowledge support by the Central 
Institute for Applied Mathematics at Research Centre J{\"u}lich. 
S.E.~is supported by the Honjo International Scholarship Foundation. 
Also acknowledged is partial support by Grants-in-Aid for Scientific 
Research from the Ministry of Education, Science, Sports, 
and Culture of Japan. 
A part of computations was carried out at the computer 
centers of the Institute for Molecular Science, Okazaki, 
and the Institute for Solid State Physics, University of 
Tokyo.

\end{document}